# Speech Emotion Detection Based on MFCC and CNN-LSTM Architecture

Qianhe Ouyang[1, *]

[1]Department of Information and Software Engineering, University of Electronic Science and Technology of China, No.2006, Xiyuan Ave, West Hi-tech Zone Chengdu, Sichuan, 611731

[*]Corresponding author email: 2020090904016@std.uestc.edu.cn

**Abstract.** Emotion detection techniques have been applied to multiple cases mainly from facial image features and vocal audio features, of which the latter aspect is disputed yet not only due to the complexity of speech audio processing but also the difficulties of extracting appropriate features. Part of the SAVEE and RAVDESS datasets are selected and combined as the dataset, containing seven sorts of common emotions (i.e. happy, neutral, sad, anger, disgust, fear, and surprise) and thousands of samples. Based on the Librosa package, this paper processes the initial audio input into waveplot and spectrum for analysis and concentrates on multiple features including MFCC as targets for feature extraction. The hybrid CNN-LSTM architecture is adopted by virtue of its strong capability to deal with sequential data and time series, which mainly consists of four convolutional layers and three long short-term memory layers. As a result, the architecture achieved an accuracy of 61.07% comprehensively for the test set, among which the detection of anger and neutral reaches a performance of 75.31% and 71.70% respectively. It can also be concluded that the classification accuracy is dependent on the properties of emotion to some extent, with frequently-used and distinct-featured emotions having less probability to be misclassified into other categories. Emotions like surprise whose meaning depends on the specific context are more likely to confuse with positive or negative emotions, and negative emotions also have a possibility to get mixed with each other.

## 1. Introduction

Emotion detection currently has become a nascent and attractive field where many researchers applied different applications, and the potential necessity and significance have gradually emerged with the increasing appearance of human-computer circumstances. The emotion detection services could enable users to have enhanced experiences to manipulate and adjust their real-time emotions, receiving feedback timely. The computer which possesses the ability to perceive and respond to human non-lexical communication including emotions could customize the settings according to certain needs and preferences. For instance, in real-life applications like Telemarketing and commercial activities, staff can also apply emotion detection techniques to develop and advance sales tactics analytically [1]. Additionally, the techniques are also applied to virtual reality situations to better simulate and evoke emotions [2]. Thus, the captured emotions can be handled and converted into a fixed sort of emotional expression in the virtual world.

Holistically, human emotions are mainly collected and conveyed from three aspects [3], namely video records, facial expressions, and vocal audio, of which both the latter two respects are separated and extracted from videos. When it comes to the facial expression research direction, multiple previous studies have already achieved profound study results from corresponding fields using different algorithms and optimization strategies [4], of which the most famous including Hidden Markov Models, Bayesian networks, EmotioNet, etc. From the vocal audio aspect, however, the emotion perception under speech cases is an open-ended problem yet. To comprehend the mindset of the human voice as much as possible, the machine is required to evaluate the person who is leading the current sentence (i.e. gender features and speaking habits) and the context of a conversation. Compared to facial expressions recorded in image forms, the human voice is more personalized and versatile, carrying a multitude of emotions. Thus, there still exists considerable uncertainty about speech emotion in the machine learning field due to the complicated feature processing steps, and what features influence the recognition of speech emotion is still a controversial topic to be discussed. During the detection processes, speech emotions are identified and classified into multiple categories, whose accuracy varies predominantly depending on a certain chosen algorithm and optimization according to context.

With respect to previous research, early in the year 2015, Hari Krishna Vydana generally adopted Gaussian mixture modeling with a universal background model (i.e. GMM-UBM) to develop a speech emotion recognition system, taking only four kinds of emotions (i.e. Anger, Fear, Happy and Neutral) into consideration [5]; later on, Ismail Shahin incorporated several factors to develop a classifier that combines a cascaded Gaussian mixture model with a deep neural network (i.e., GMM-DNN) and performs better in the presence of noisy signals [6]. However, there still exists certain flaws such as the high training difficulty and low efficiency when dealing with large dataset for the Gaussian mixture modeling approach. Also, for the GMM-DNN method, it would be quite demanding to record changes of samples based on chronological sequences which is vital for speech processing scenes.

In this paper, CNN-LSTM architecture is implemented in order that the most conspicuous and featured properties could be extracted from massive input and the characteristic of feedback connections over time can be inherited at the same time. Part of the mainstream and widely-selected vocal audio processing techniques including Formants of speech and Mel Frequency Cepstral Coefficients (MFCCs) are adopted in the paper for the sake of guaranteeing the integrity of the input to the classification algorithms as well as the efficiency. The captured speech signals are recorded in the form of time domain plot thus creating accurate image input to analyze for the CNN-LSTM model. Besides, up to seven sorts of speech emotions (happy, neutral, sad, anger, disgust, fear, surprise) are extensively trained and categorized, enlarging the potential prospects of the speech emotion detection system with a more precise and accurate service. These numerous sorts of classifications are able to explore and explain which speech emotion carries the most information and contributes to the underlying reasons. The research result also could be regarded as a criterion of emotion classifications comprehensively to some extent.

The rest of this paper is organized as follows: Section 2 describes the theoretical techniques from the original audio processing to feature extraction to the architecture of CNN-LSTM neural networks. In Section 3, the final results and performances applying this model are discussed from multiple metrics thoroughly. The final conclusions of this work and further potential enhancement in the future are discussed in Section 4.

## 2. Methods

*2.1. Dataset description and preprocessing*
For the dataset, two relatively common-used datasets are taken into consideration as the original speech signal. The Surrey Audio-Visual Expressed Emotion (SAVEE) dataset was recorded from four male native English speakers (identified as DC, JE, JK, KL) who are also postgraduate students aged from 27 to 31 years [7]. Emotion in the dataset has been described psychologically in seven discrete categories: neutral, anger, disgust, fear, happiness, sadness, and surprise. For each emotion per speaker, there are 3 common, 2 emotion-specific, and 10 generic sentences which are different for each emotion and phonetically-balanced, thus totally constituting 480 utterances. All of the utterances in the Ryerson Audio-Visual Database of Emotional Speech and Song (RAVDESS) collection [8] are gender balanced and are voiced by 24 professional actors with a neutral North American accent. Only a portion of the speech section from the dataset—more than 2000 vocal audio files—was chosen. This portion contains expressions for calm, happy, sad, furious, afraid, surprise, and disgust, created at two different levels of emotional intensity, as well as a neutral expression.

Combining the two datasets mentioned above, a specific and customized dataset could be created with seven distinct emotions: happy, neutral, sad, anger, disgust, fear, and surprise. The dataset has 2, 459 samples in total with around 350 samples on average for each emotion respectively. Incidentally, the speech audios in calm emotion are not adopted and removed in this dataset due to the unsatisfactory training result and the relatively larger possibility of confusion with other emotions which cannot be underestimated.

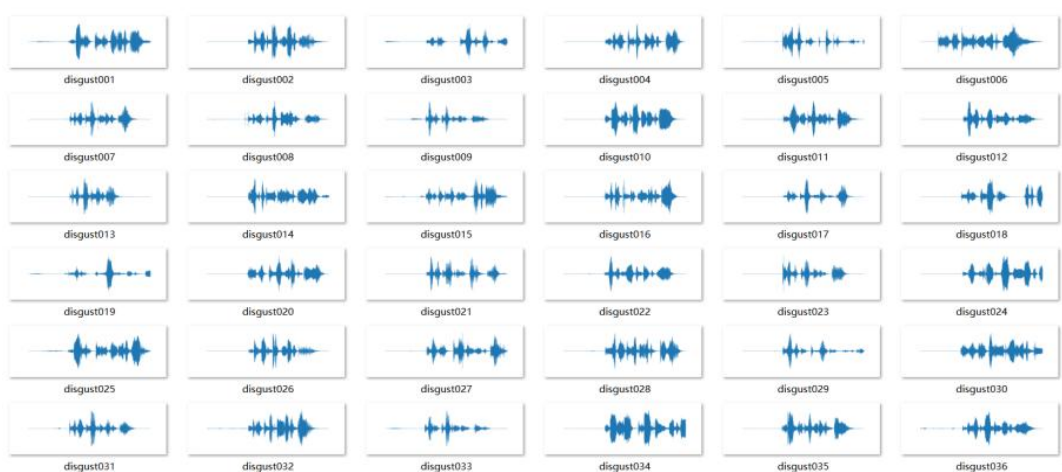

**Figure 1.** the waveplot of audios

In regard to the data preprocessing procedure, every single audio file is supposed to be converted into an image file as shown in Figure 1 to create the initial input for the convolutional neural network. Also, each file has many labels as identifier, basically including the emotion label and gender label. To reduce the complexity of time-domain sound representation, separation of male and female voices is executed as the experimental accuracy increased by 10% or so, reducing the workload of pitch analysis. Overall, the converted image files still provide various features appended by the labels and recorded in the array forms. Thus, the Librosa package—a popular Python package for audio and music signal processing that offers implementations of a variety of common functions—is utilized to carry out the major procedures of feature extraction [9]. Measuring the average energy of signals is contemplated as an approach to excavating the key characteristics, i.e. the volume of the audio. Parameters related to the whole signal duration like the mean value, standard deviation, maximum, and minimum also attach significance because they indicate the pitch and emotional fluctuations which are decisive to emotion classification. Skewness and kurtosis are specifically derived for the analysis of the signal plots.

The Fast Fourier Transform (FFT) is a vital procedure in the whole MFCC process, it processes the signal in the frequency domain to serve the subsequent procedures of Mel Frequency Warping. The power spectrum can be derived as a result of FFT by taking the square value of the signal at windowed samples which record the frequency content of the signal at different points. The values of the power spectrum are saved in the feature vector, and the three largest frequency peaks for each window are additionally recorded as features as they always represent the greatest power. The maximum and minimum frequencies are used to determine the frequency range of each signal frame, combing the FFT bins into equally spaced intervals. Therefore, the auditory spectrum could eventually be created by mapping the power spectrum mentioned above to an auditory frequency axis.

The MFCC can be derived by mapping the powers of the frequency spectrum onto the Mel scale and discrete cosine transformation. The Mel scale is based on a mapping between actual frequency and perceived pitch because it appears that the human auditory system does not perceive pitch in a linear manner [10].

There are two scales used to record alterations of pitch over time, namely the coarse time scale and the fine time scale. For the previous one, the signal is simply divided into the beginning, middle, and ending parts, of which the part with the highest average pitch is employed to pitch rising (falling) detection. For the latter scale, comparisons are made between the dominant frequency of each windowed sample and its preceding and following one. And such discrepancy in each sample is also recorded as features in the vector.

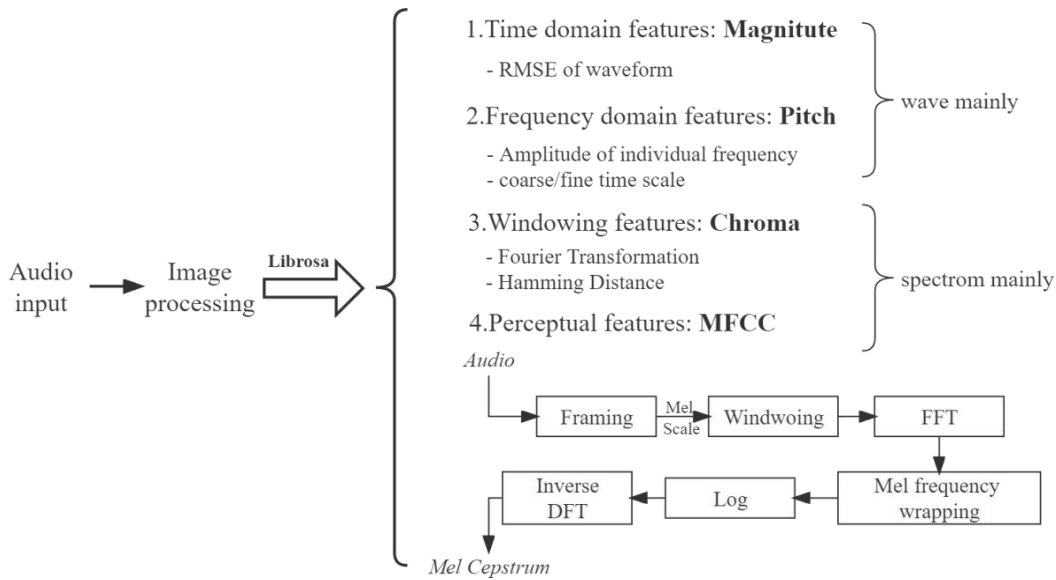

**Figure 2.** Feature extraction procedures

Above all, after executing the feature extraction process following the principles and parameters mentioned above, countless features are gained through the overall method as shown in Figure 2. However, a further step of feature selection is required, which can help filter and avoid the high variance of the algorithm as well as overfitting. For each single audio sample, the speech signal has roughly 70 windows, and each windowed sample contributes about 570 features in total. Thus, heuristics are implemented to score each feature to select the most representative features from more than 40,000 features altogether, instead of using brute force forward or backward search.

*2.2. CNN-LSTM architecture*
The hybrid CNN-LSTM model inherits the characteristics of both CNN and LSTM, the layers of the former network are mainly used to extract features from input data, while the latter network has the capability to tackle perceptual problems related to the time-varying scenarios. It belongs to a category of typical models that are deep in both space and time and adaptable enough to be used for a range of vision tasks involving sequential inputs and outputs [11].

As illustrated in Figure 3, the holistic structure mainly can be described in the following steps:
I The Convolutional Layer includes filters that extract features to pass on to the following layer, and the convoluted output is regarded as an activation map. Zero padding is used to ensure that the filter is applied to all the elements of the input, enabling further computation to be more advantageous.

II The Dropout Layer is designed in order to prevent complex co-adaptations and overfitting on the training data though the training time might be greatly extended.

III The Max Pooling Layer is adopted to speed up computation for upper layers for removing non-maximal data.

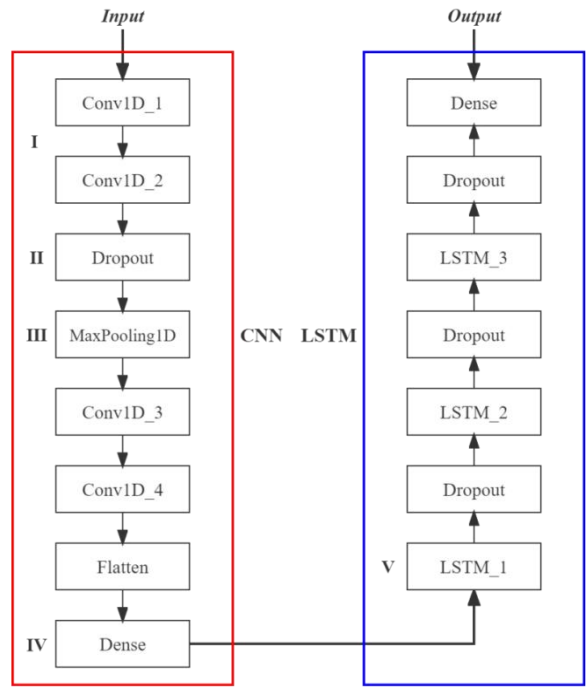

**Figure 3.** The architecture of CNN-LSTM

IV The Dense Layer is a fully connected layer where the flattened input is processed into the classification output. Afterward, the output is contrasted to verify errors through the loss function.

V With respect to the LSTM module, 3 LSTM layers are implemented to enhance the robustness to deal with more complex sequences and unstable time series that may exist in partial samples.

*2.3. Implementation details*

The architecture is implemented in Tensorflow, and the settings of parameters are shown as follow: RMSprop is adopted as the optimizer, with a 0.00001 learning rate and 1e-6 decay. Same padding is used for the feature map. Activation and loss function are relu and categorical_crossentropy respectively. The dropout value for the CNN and LSTM part is 0.1 and 0.2 correspondingly. With respect to the training settings, 370 epochs are taken with a batch size of 16, and the overall evaluation metrics is accuracy.

## 3. Result and discussion

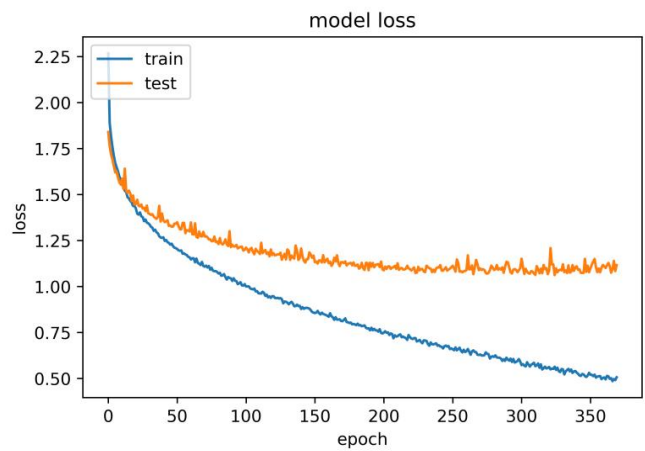

**Figure 4.** The model loss for training and test set

Firstly, the basic training effects of the model using the architectures designed above are illustrated in Figure 4. It could be illustrated in Fig.4 that the loss of the model decreases in the whole process with the increase of epochs for the training set and continues to decline with the trend of getting closer to zero, while the loss for the test set tends towards stability after the epochs reaching approximately 200. The probability of overfitting is reduced as much as possible, and the settings of parameters are appropriate so that the whole training effect is relatively ideal according to the loss.

**Table 1.** Detection accuracy of different emotions

| Classification | anger | disgust | fear | happy | neutral | sad | surprise |
|---|---|---|---|---|---|---|---|
| **Mean Accuracy** | 75.31% | 38.33% | 59.04% | 61.18% | 71.70% | 56.70% | 56.60% |

The average accuracy of the speech emotion recognition in different classifications are demonstrated in Table 1. The accuracy is calculated averagely after the training process for each sort of emotion. Overall, the classification accuracy reaches 61.07% comprehensively. The accuracy of anger and neutral are the highest among the seven emotions at above 70%, while the classification effect of disgust performs the worst at merely 38.33%. According to relative research, the presence of overlapping and emotions with similar features in speech has made recognition in this field intriguing and complex [12]. Genuinely, the anger emotion which has the highest accuracy at 75.31% is regarded to have a large proportion of similarity with the disgust and sad emotion, though negative emotions do have distinct prosodic features. Hence, the low accuracy of disgust emotion is considered to be reasonable due to the insignificant differences.

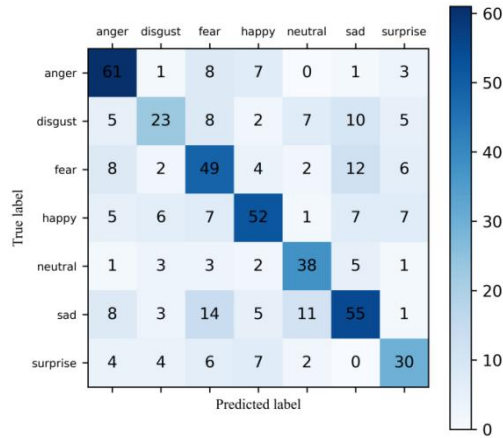

**Figure 5.** The confusion matrix of the results

From the confusion matrix as illustrated in Figure 5, the recognition effect could be observed more clearly by analyzing part of the misclassification situations. For the negative emotions like sad, it is likely to be classified into fear and neutral emotion with 14 and 11 samples respectively, while the fear emotion has certain possibility to be classified into the sad emotion with 12 cases. Similar mixed circumstances can be discovered in the misclassification between surprise and fear, surprise and happy, which is probably due to the pluralistic definition of surprise based on positive or negative context of a specific speech. Common-use emotions like anger, happy, and neutral perform fewer misclassification cases compared to the negative emotions, which enables the overall performance of the speech recognition system to be tolerable and satisfactory to some extent.

## 4. Conclusion
In this paper, the problem of emotion detection in speech cases is tackled using the designed feature extraction methods and the CNN-LSTM network as mentioned above. Regarding the feature extraction, features from four aspects are adopted for further analysis and the MFCC processing provided by the Librosa is specifically vital. After the preprocessing stage, the customized CNN-

LSTM architecture accomplishes a decent training effect with an overall 61.07% accuracy for 7 categories. Results illustrated in the confusion matrix indicate that the emotion detection accuracy is dependent on the properties of emotions as the distinct-featured emotions like happy and anger could be classified more precisely, while negative emotions like sad, fear, and disgust are more likely to get mixed with each other. Further enhancement of the detection could focus on improving the accuracy for negative emotions by implementing models whose structures are more complicated. Besides, the detection system could be further applied to robots or software after integration, enabling the machines to have the potential to communicate and converse just as people do.

## 5. References

[1] Ruiz L Z et al. 2017 Human emotion detection through facial expressions for commercial analysis 2017IEEE 9th International Conference on Humanoid, Nanotechnology, Information Technology, Communication and Control, Environment and Management (HNICEM) pp 1-6
[2] Felnhofer A et al. 2015 Is virtual reality emotionally arousing? Investigating five emotion inducing virtual park scenarios International journal of human-computer studies 82 pp 48-56
[3] Yu J 2014 A video, text, and speech-driven realistic 3-D virtual head for human machine interface IEEE transactions on cybernetics 45(5) pp 991-1002
[4] Adeyanju I A et al. 2015 Performance evaluation of different support vector machine kernels for face emotion recognition 2015 SAI Intelligent Systems Conference (IntelliSys) pp 804-806
[5] Vydana H K et al. 2015 Improved emotion recognition using GMM-UBMs 2015 International Conference on Signal Processing and Communication Engineering Systems pp 53-57
[6] Shahin I et al. 2019 Emotion recognition using hybrid Gaussian mixture model and deep neural network IEEE access 7 pp 26777-26787
[7]Eu J L 2019 Surrey Audio-Visual Expressed Emotion 2022 https://www.kaggle.com/datasets/ejlok1/surrey-audiovisual-expressed- emotion-savee
[8]Steven R L 2018 RAVDESS Emotional speech audio 2022 https://www.kaggle.com/datasets/uwrfkaggler/ravdess-emotional-speech- audio
[9] McFee B et al. 2015 librosa: Audio and music signal analysis in python Proceedings of the 14th python in science conference Vol 8 pp 18-25
[10] Logan B 2000 Mel frequency cepstral coefficients for music modeling International Symposium on Music Information Retrieval
[11] Donahue J et al. 2015 Long-term recurrent convolutional networks for visual recognition and description Proceedings of the IEEE conference on computer vision and pattern recognition pp 2625- 2634
[12] Palo H K et al. 2015 Classification of emotions of angry and disgust SmartCR 5(3) pp 151-158